# Unveiling the Potential of Big Data Analytics for Smart Education in Bangladesh: Needs, Prospects, Challenges, and Implications

Sabbir Ahmed Chowdhury[1,2], Md Aminul Islam[3*], Mostafa Azad Kamal[4]

[1]School of Education and Social Sciences, University of the West of Scotland, UK

[2] Institute of Education & Research, University of Dhaka, Bangladesh

[3] School of Engineering, Computing and Mathematics, Oxford Brookes University, UK

[4] School of Business, Bangladesh Open University, Gazipur

*Corresponding Author: mdaminulislam1@connect.glos.ac.uk

**Abstract**:

Big Data Analytics has gained tremendous momentum in many sectors worldwide. Big Data has substantial influence in the field of Learning Analytics that may allow academic institutions to better understand the learners' needs and proactively address them. Hence, it is essential to understand Big Data and its application. With the capability of Big Data to find a broad understanding of the scientific decision-making process, Big Data Analytics (BDA) can be a piece of the answer to accomplishing Bangladesh's Higher Education (BHE) objectives. This paper reviews the capacity of BDA, considers possible applications in BHE, gives an insight into how to improve the quality of education or uncover additional values from the data generated by educational institutions, and lastly, identifies needs and difficulties, opportunities, and some frameworks to probable implications about the BDA in BHE sector.

**Keywords**: Big Data Analytics, Learning Analytics, Quality of Education, Challenges, Higher Education, Bangladesh.

1. **Introduction:**

The global landscape is undergoing rapid transformation due to the advent of innovative technologies (Mention, 2019). Presently, there is a considerable proliferation of technical equipment being utilized by individuals (Shorfuzzaman et al., 2019) and a substantial volume of data is generated continuously by these devices (Baig, Shuib and Yadegaridehkordi, 2020). According to Kalaian, Kasim, and Kasim (2019), utilizing these technologies and applications proves advantageous in data analysis and storage. As a result, the term "Big Data" has gained significant attention and popularity in several domains such as business, education, health studies, statistics, and numerous other disciplines (Liang and Liu, 2018). The concept of big data is defined by three core dimensions, commonly known as the three Vs, which include Volume, Velocity, and Variety. The quick and extensive proliferation of data across many platforms and sources, such as mobile devices, social media platforms, business transactions, and personal travel patterns, has emerged as a significant issue due to its substantial scale and accelerated rate of expansion (De Mauro, Greco, and Grimaldi, 2016). The concept of the three V's has been broadened to include several supplementary V's. Demchenko, Grosso, De Laat, and Membrey (2013) introduced a classification paradigm for big data, which has five fundamental dimensions widely known as the 5Vs. The dimensions encompassed in this framework consist of Volume, Velocity, Variety, Veracity, and Value. In their publication, Saggi and Jain (2018) have provided a thorough explanation of big data, encompassing seven distinct aspects that are popularly referred to as the 7 V's. The dimensions being considered involve Volume, which relates to the significant amount of data generated; Velocity, which relates to the speed at which data is generated and processed; and Variety, which encompasses the diverse range of data types and sources.

When conducting data analysis, it is typical to evaluate four dimensions: valence, honesty, variability, and value. Valence refers to the inherent positive or negative emotional connotation associated with information. The notion of veracity pertains to the accuracy and reliability of the data. Variability refers to the presence of irregularity and volatility that is inherent within a given dataset. The term "value" pertains to the intrinsic capacity of data to possess utility and significance. There has been a significant increase in the demand for big data in several areas, such as insurance and construction, healthcare, communications, and e-commerce (Baig, Shuib, & Yadegaridehkordi, 2020).

In recent years, a notable transformation has been observed in the domain of Higher Education (HE) across many nations. The term "transition" pertains to the incorporation of novel digital data sources that are being produced, examined, and employed to enhance decision-making within higher education (Williamson, 2018). The education domain generates a substantial amount of data because of the widespread availability of online courses and diverse teaching and learning practices (Yamada et al., 2017). According to Lovett and Wagner (2012), implementing Big Data Analytics (BDA) enables students to monitor their advancement in both academic and behavioral spheres proficiently. Furthermore, this training enhances the instructor's ability to observe and assess student performance.

According to Aslam and Khan (2020), educators can gather vital data on students' academic accomplishments and learning tendencies, enabling them to deliver feedback promptly. Yu and Gao (2022) argue that timely and constructive feedback to students positively affects their motivation and satisfaction, ultimately leading to improved academic performance. The data indicated above can be employed for enhancing pedagogical approaches, influencing the creation of curricular materials, and tailoring educational experiences for individual students (Williamson, 2018). In addition, the emergence of numerous online educational platforms has permitted the development of diverse courses tailored to individual students' distinct interests (Maghsudi et al., 2021). The enhancement of the educational sector is contingent upon the acquisition and integration of technology. Using extensive administrative data holds significant potential in effectively addressing academic challenges (Baig, Shuib, & Yadegaridehkordi, 2020). Hence, professionals must comprehend the efficacy of big data in education to mitigate educational challenges. The scientific and technological advancement and economic progress of a nation heavily depend on the capabilities of its higher education system. Higher education programs focus on connecting big data and long-term objectives and goals. Therefore, it is imperative to investigate the impact of big data analytics on the learning paradigm and its role in enabling the transition to technology-mediated learning in developing nations. This chapter delineates the ramifications and obstacles associated with implementing BDA in educational contexts inside Bangladesh.

2. **Conceptual Framework**:

The phenomenon of Big Data is experiencing significant expansion and has already attained extensive implementation across diverse sectors, including healthcare, business, and government. As a nascent field of inquiry within postsecondary education, it comprises research areas such as Educational Data Mining (EDM) and learning analytics (LA). The primary focus of EDM is the promotion and development of computational technologies that enhance the process of identifying patterns within educational data. In contrast, the focus of LA is directed at understanding the academic progress of individual students within a particular educational environment (Waheed et al., 2018). When viewed through a theoretical lens, the notion of Big Data in education pertains to vast and significant amounts of educational data, regardless of whether they exist in physical or digital formats and are housed in diverse repositories. These repositories consist of various sources, encompassing tangible documents such as account books that educational institutions meticulously maintain. Additionally, they include records about class tests, examinations, and the alumni of these schools. According to Bharathi (2017), to successfully deploy Big Data solutions within the context of higher education, it is imperative to precisely understand and comprehend diverse sets of administrative and operational data. The data in question is important when it comes to assessing performance and progress and identifying potential obstacles related to academic programs, research, teaching, and learning (Long and Siemens, 2014; Picciano, 2012). Daniel and Butson (2013) developed a conceptual framework to conceptualize the concept of Big Data in the context of higher education. The framework has four components, specifically institutional analytics, information technology analytics, learning analytics, and academic analytics (Figure 1).

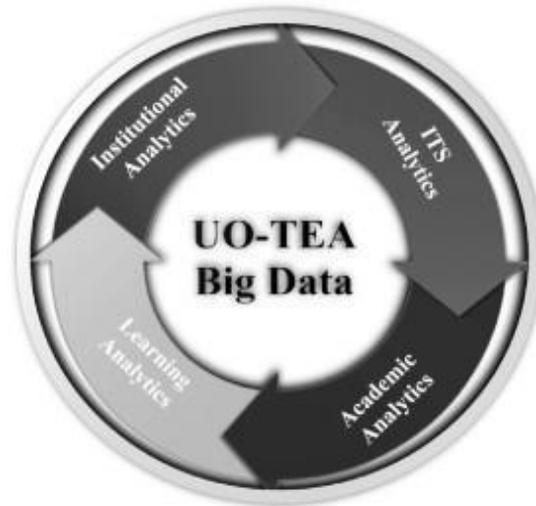

Figure 1: Conceptual Framework [Adopted from Daniel and Butson (2013)]

- **Institutional Analytics**: The term Institutional Analytics (IA) encompasses a range of operational data that can be analyzed to facilitate informed decision-making to enhance institutional performance (De Silva et al., 2022). Information Architecture (IA) encompasses the analysis of assessment policies and structural elements. Institutional analytics utilize tools such as reports, data warehouses, and data dashboards to enable institutions to make informed decisions based on timely and data-driven insights (Daniel, 2015).
- **Information Technology Analytics:** The field of Information Technology Analytics (ITA) encompasses the collection and analysis of usage and performance data, which is crucial for monitoring and improving technology development, data standards, tools, procedures, and regulations. IT analytics aims to consolidate and incorporate all gathered data (Daniel, 2015).
- **Academic Analytics**: Academic Analytics (AA) incorporates all activities within higher education that impact administration, research, resource allocation, and management (Quadir et al., 2022). AA pertains to data analysis at an institutional level while learning analytics focuses on examining the learning process, encompassing the study of the interplay between the learner, content, institutions, and educator (Yassine, Kadry, and Sicilia, 2016).
- **Learning Analytics**: The field of Learning Analytics (LA) focuses on the systematic gathering, examination, and presentation of data about learners and their specific learning environments. The primary objective of LA is to gain insights and enhance the learning process, as well as the various settings in which it takes place (Siemen, 2012). Marks, Al-Ali and Rietsema (2016) assert that the utilization of learning analytics software and approaches is prevalent to enhance processes and workflows, assess academic and institutional data, and enhance the overall effectiveness of organizations.

### 3. Educational Evolution in response to Emerging Technologies

Education 1.0 was based on traditional approaches, where teachers played a central role in imparting knowledge. The emergence of Education 2.0 can be attributed to the advancement of technology, specifically the integration of electricity and mass manufacturing. This shift resulted in a reconceptualization of the responsibilities of educators, situating them as disseminators of knowledge. Education 3.0 integrates automation and internet resources, facilitating a conducive environment for collaboration between instructors and learners. Within the realm of Education 4.0, there are prominent advancements in technology, including artificial intelligence (AI), data management systems, and robotics. There is a growing trend among nations and academic institutions to embrace Education 4.0, highlighting the importance for educators to develop proficiency in these areas (Ahaidous et al., 2023).

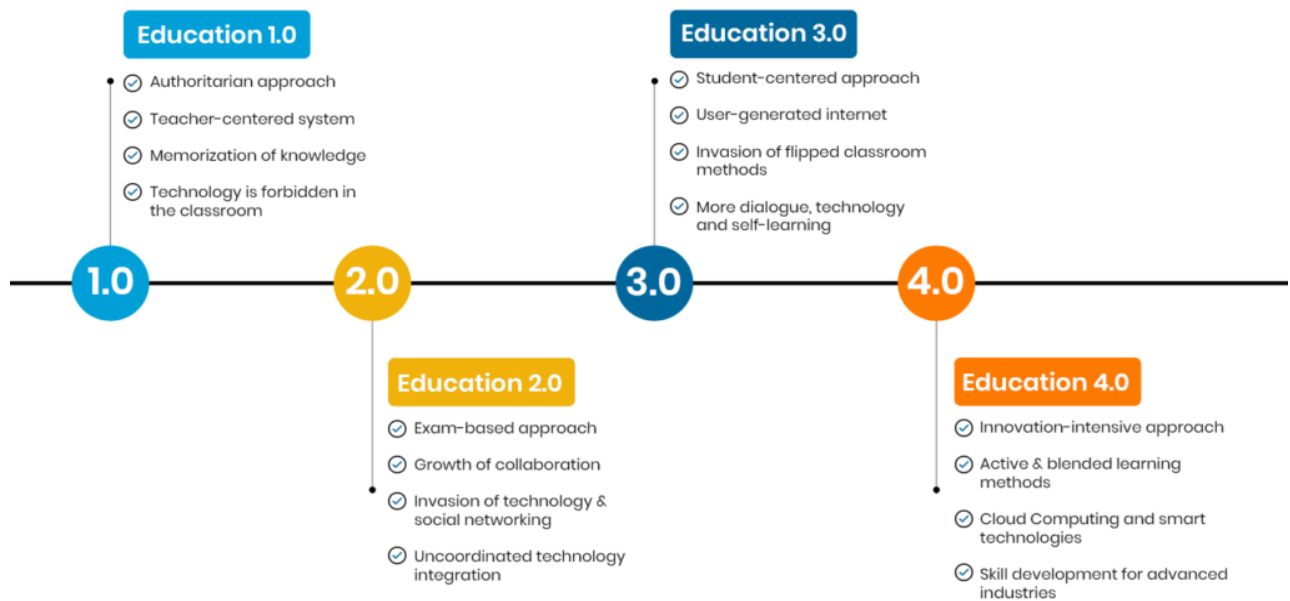

Figure 2. Evolution of Education 4.0

### 4. Approaches of Big Data Analytics

According to Daniel (2015), the utilization of big data and analytics in higher education institutions can offer predictive capabilities that can enhance individual student learning outcomes and assure the maintenance of high-quality academic programs. The utilization of big data and analytics in the context of higher education has the potential to bring about significant transformations in various aspects, including the administration process, teaching methodologies, learning outcomes, and scholarly endeavors (Baer & Campbell, 2011). Although there is considerable excitement about different approaches to BDA, applying these methods can be demanding regarding manpower. This study emphasized the utilization of many methodologies, including widely available software tools such as Hadoop and MapReduce, the extension of pre-existing software, and the introduction of innovative approaches like DEMass for density estimation. The methodologies described involve various analytical procedures, such as text analysis, audio analysis, video analysis, social media analysis, predictive analytics, descriptive analytics, inquisitive analytics, prescriptive analytics, and pre-emptive analytics. The systematic literature review (SLR) categorized analytics methods into three distinct classes, namely descriptive, predictive, and prescriptive analytics (Sivarajah et al., 2017).

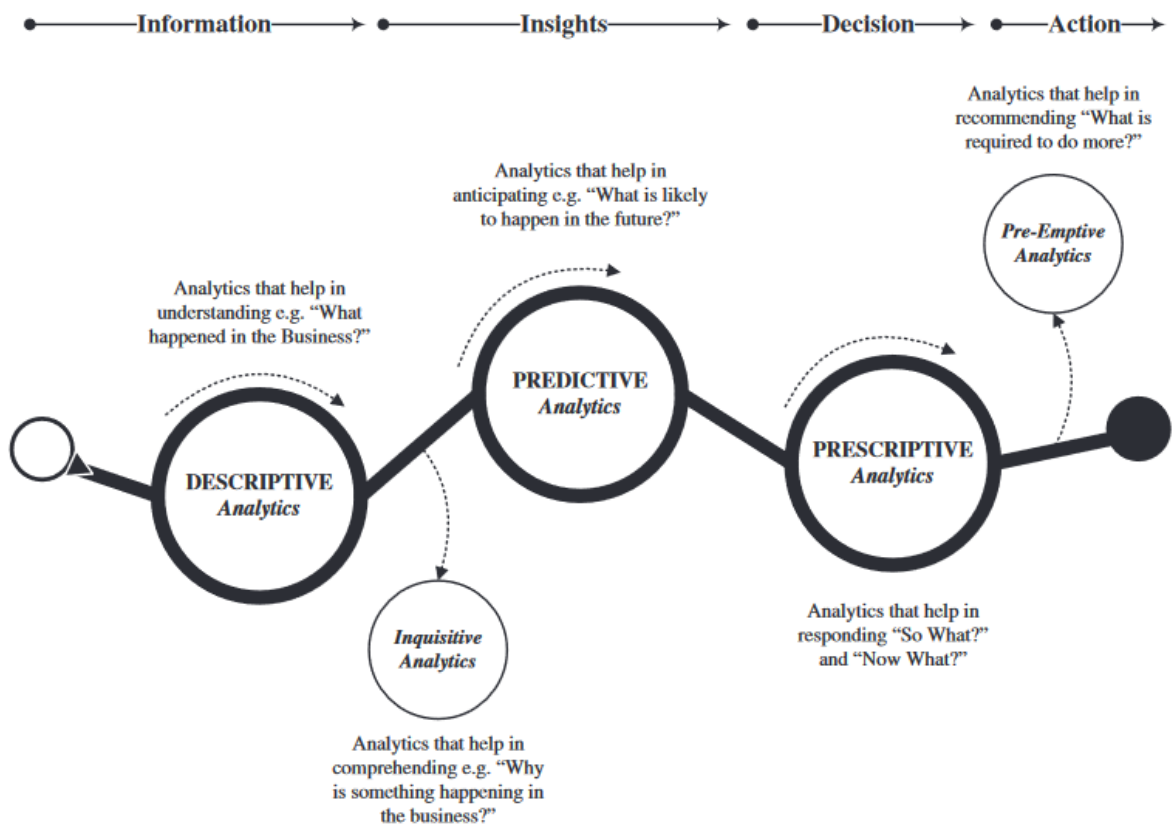

Figure 3. Types of BDA

Like other businesses, descriptive, predictive, and prescriptive analytics can be used for higher education (Sivarajah, U., et al., 2017) which has been articulated in Fig 3. Descriptive analytics has the potential to be utilized to:

-Examine the characteristics of student populations and analyze patterns in enrollment. This data can be utilized to make well-informed decisions regarding allocating resources, scheduling courses, and implementing marketing activities.

- Monitor student academic progress. This data can be utilized to ascertain students encountering difficulties and offer them supplementary assistance. Assess the efficacy of various pedagogical approaches and initiatives. The provided information possesses the potential to inform data-driven decision-making regarding curriculum creation and instructional practices.

Predictive analytics can be utilized for:

-This study aims to forecast student attrition rates within an academic institution. The provided data can be utilized to formulate early intervention initiatives to assist students with a heightened risk of discontinuing their education. The objective is to forecast the academic performance of students in particular courses. This data can offer students individualized learning trajectories and assistance services.

-Determine the demographic of students more likely to derive advantages from forms of financial assistance. This data can enhance the precision of financial aid resource allocation.

Prescriptive analytics can be utilized for:

- propose the implementation of individualized learning trajectories for pupils. This information can assist students in making well-informed decisions regarding their academic pursuits.

-Propose course schedules that are strategically designed to enhance student achievement. This information can aid in ensuring that students are enrolling in appropriate courses at the correct times.

-Determine the students who are susceptible to making suboptimal academic choices. This data can be utilized to implement early interventions and offer pupils the necessary assistance to improve their decision-making abilities.

The higher and professional domain necessitates ongoing evaluation and adaptation to keep up with evolving trends across many market sectors. Consequently, this dynamic environment gives rise to many labor requirements (Deepa & Chandra, 2017). According to proponents, the use of big data analytics in the field of education is expected to generate several prospects for educational institutions, administrators, policymakers, educationalists, and learners. One potential benefit is facilitating collaboration and comparison among the institutions. According to Wagner and Ice (2012), the use of collaborative ventures in big data efforts can contribute to the collective ownership of difficulties related to student performance and persistence. The organization would realize enhanced knowledge dissemination and increased learning outcomes. This aligns with the perspective that the significance of big data lies in its capacity to facilitate the collaborative establishment of governing frameworks and the implementation of more advanced and superior policies and strategies that are now employed (Schleicher, 2013). According to Mukthar et al. (2017), the self-measurement of both learners and educators has the potential to enhance learning effectiveness. The potential for cost reduction exists through the management of financial performance. It is possible to reduce the challenges and complexity associated with learning and academic endeavors.

### 5. Challenges of Big Data

Despite having many necessities and opportunities, several anticipated challenges are associated with implementing analytic techniques for BDA in higher education (HE).

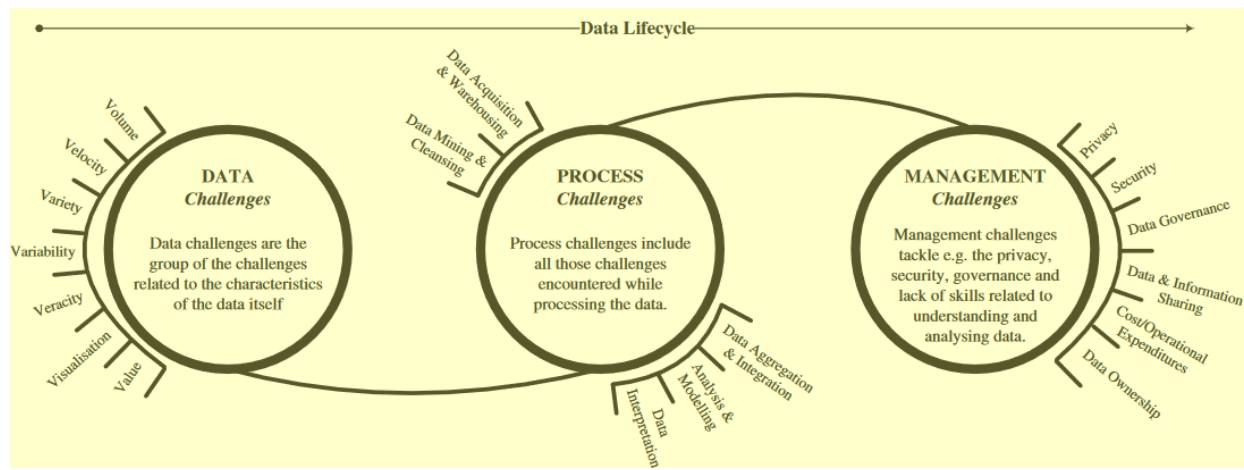

Figure 4: Challenges of Big Data

The issues associated with Big Data can be classified into three distinct categories: data, process, and management, as illustrated in Figure 4. Data difficulties encompass various factors, including but not limited to volume, velocity, variety, variability, veracity, visualization, and value. The obstacles encountered in the process encompass many stages such as Data Application and Warehousing, Mining and Cleansing, Aggregation and Integration, Analysis and Modelling, and Interpretation. The field of management presents several issues, including those related to privacy, security, governance, data and information sharing, cost and operational expenditures, and ownership. Ensuring data availability for analysis poses a significant obstacle in the implementation of educational analytics (Deepa & Chandra, 2017). Higher education institutions are faced with the expectation of meeting various demands from different stakeholders. However, this expectation comes at a time when government funding is decreasing, support from corporate and private sectors is dropping, and there is an increasing need for clarity and responsibility due to regulatory expectations (Hazelkorn, 2007). Obtaining necessary data from the weakly integrated database system poses challenges, as does the development of a comprehensive data warehouse for all institutions (Shikha, 2014). Moreover, data integration difficulties are prominent, particularly when data is available in both structured and unstructured formats and necessitates integration from several sources, many of which are housed in systems overseen by separate departments (Daniel, 2015). According to Deepa and Chandra, the presence of low-quality and improperly formatted data originating from a less easily accessible database system can give rise to substantial issues. According to Dringus (2012), it is recommended that learning analytics be designed with transparency and flexibility to enhance accessibility for educators. Gaining a more comprehensive comprehension of the practitioners of the system will require a significant investment of time. The task of conveying information in a manner that is both comprehensible and enlightening poses a significant challenge for both learners and educators (Shikha, 2014). Collaboration can be hindered by a persistent divide between individuals possessing expertise in data extraction and knowledge of existing data, and those possessing expertise in identifying necessary data and determining optimal utilisation methods (Daniel, 2015). According to Romero and Ventura (2010), generating relevant context for analytics has historically been challenging for those without specialized expertise. This difficulty has hindered the capacity of non-specialists to effectively visualize data in compelling ways and comprehend its significance, hence reducing the observability and impact of analytics. Merely implementing the analytical system does not suffice to surmount all the hurdles. Facilitating practitioners' familiarity with the design and fostering their willingness to collaborate would be a considerable challenge. Consequently, if the utilization of analytics is not effectively guaranteed, all endeavors will be rendered futile (Deepa & Chandra, n.d.). Deficiencies in risks and security procedures pertaining to data protection and privacy persist at numerous higher education institutions. According to Slade and Prinsloo (2013), most higher education institutions have implemented policies to manage intellectual property, protect data privacy, and regulate data access. However, these policies may not be sufficient in addressing the current challenges posed by big data in the higher education sector. Big Data involves the collection of a substantial volume of data, sourced from many channels within the institutional database (Picciano, 2012). The emergence of privacy concerns among students, teachers, and staff necessitates that institutions adopt appropriate measures to effectively tackle ethical and privacy-related challenges (Kalota, 2015).

Various expected intricacies are identified with implementing big data analytic techniques in Bangladesh's higher education. These difficulties lie in abilities, culture, and government. Notwithstanding many needs and opportunities, educationalists must confront a few challenges in actualizing Big Data Analytics. Some of these difficulties are (Rahman et al., 2019).

1. **Acceptance of big data:**
   It includes persuading the clients to accept BDA as a path for supporting new procedures and change management because it takes more time to develop a new system and to understand it at universities in Bangladesh.
2. **Complexities in accessing data**:
   Because of the low quality and inaccurately organized data, getting the required data from the database system at Universities in Bangladesh is difficult.
3. **Needed skillful knowledge**:

Expert knowledge is required to ensure whether the system works properly or not. However, universities in Bangladesh have a shortage of experts and knowledge of Big Data.

4. **Universities barrier:** Our university authorities are not very interested in implementing Big Data, and they are worried about the cost associated with collecting, storing, analyzing, and visualizing the data.

5. **Security issues:** Security is one of the most critical issues to be considered when implementing Big Data in higher education for authentication in Bangladesh universities. These issues are related to storage, management, and processing. The introduction of BigData from various sources places an additional weight on sorting,
   Preparing and communicating. These security concerns originated from external data sources, for example, social media, which represents an uncontrolled aggregation of data and the most significant threat to data security regarding accessibility, accuracy, and dissemination of data. Therefore, the traditional data handling system is not helpful to handle Big Data in terms of security.

6. **Preserving privacy:** Protecting Privacy is another critical worry about using Big Data in higher education authentication in Bangladesh universities that needs to be considered. Privacy in Big Data for higher education in Bangladesh involves many matters such as insurance of individual information of clients, intellectual property rights, and government reports that might be spilled during data obtaining and storage. Big data must be handled by laws and regulations that protect data privacy.

7. **IT infrastructure:** Big Data creates new difficulties for both equipment and programming in the universities, essential to creating, gathering, storing, and analyzing data. Analyzing Big Data requires high storage capacity, high web speed, HPC frameworks, and the computing power to deal with the data, analyses, and client questions. So, it would be challenging to implement Big Data Analytics in higher education in Bangladesh.

## 6 Implication of Big Data in the Context of the Education System of Bangladesh

Big Data will create various opportunities for the higher education sector in Bangladesh, like educational institutions, policymakers, educationalists, and students. These opportunities incorporate (Sarker et al., 2018; Prodhan, 2016):

1. Minimizing the gap between the outcome of higher education in Bangladesh and the prerequisites of the job market and the private sector.
2. Helping students to choose their profession very cautiously.
3. Helping to improve the University rankings in Bangladesh.
4. The following improvement and publication of a refined scope of education outcomes demonstrated yearly upgrades.
5. Building a database for the students and following them from early childhood to higher education to improve education planning, monitoring, assessment, and outcomes.

6. Big Data provides a chance for Higher education in Bangladesh to utilize their data technology resources deliberately to develop education quality, assist the students to higher achievement rates, and compare with student constancy and results.

7. Any university in Bangladesh has a large amount of data, including recruitment, employees, punishing records, and so on. Analyzing and managing such data can help students and teachers find their success and areas of weakness and compare with other universities in Bangladesh.

8. The student and teachers can trace their behavioral changes using Big Data Analytics. Also, Big Data Analytics can improve the effective learning process through the self-improvement of students, and teachers, cost reduction by managing financial performance, and improving the number of graduation rates.

The successful implication of big data analytics by overcoming the challenges will lead to Smart education which has been illustrated in the figure below.

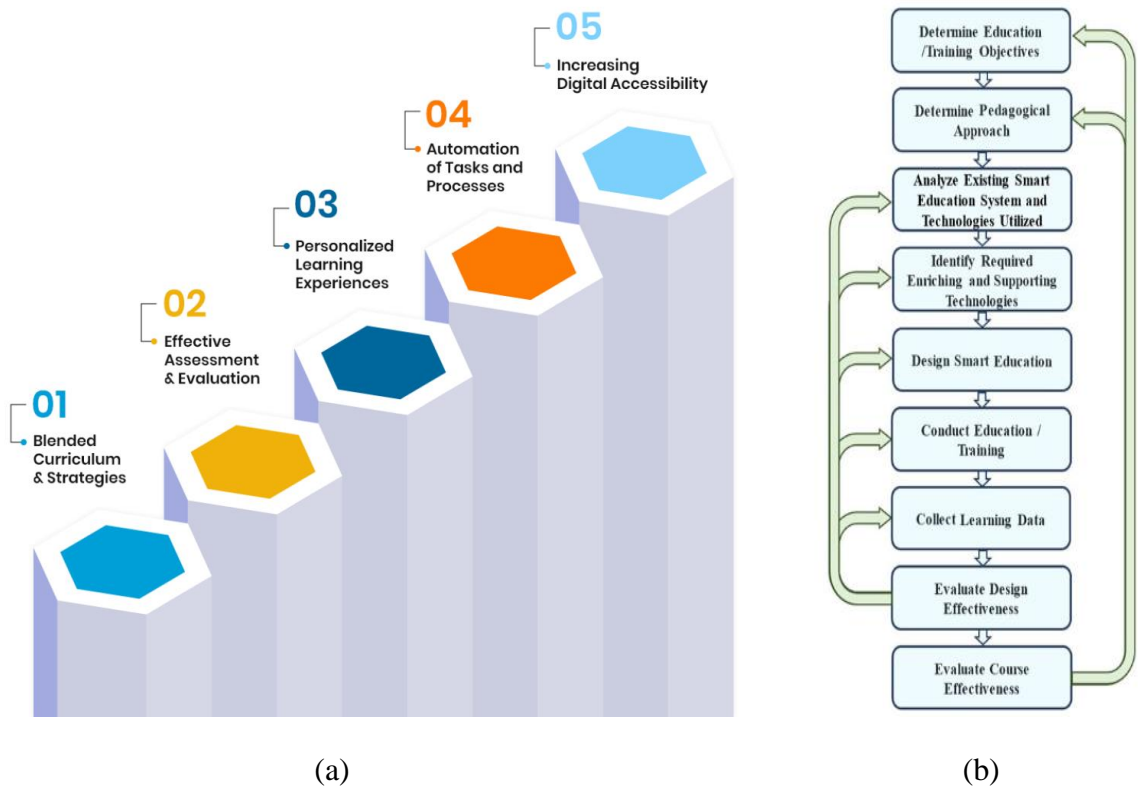

(a)                                                     (b)

Figure 5: (a) Essential Elements and (b) Design of Smart Education

Smart Education (SE) has 5 key elements which is in fig 5. BDA is a must to achieve SE which sets the objective of establishing intelligent learning environments by using innovative technologies. The aim is to facilitate tailored educational services, empower learners, and cultivate persons with heightened values, analytical reasoning, and robust behavioral competencies. Smart education has three fundamental components: smart settings, smart pedagogy, and smart learners. The framework places significant emphasis on the necessity of methodological improvements in smart pedagogy and technological upgrades in smart learning environments to foster the development of intelligent learners. This statement underscores the interconnectedness of these components, wherein intelligent instructional methods and conducive learning environments significantly impact the growth and progression of intellectually adept students (Zhu, Z.T., et., al., 2016).

BDA can bring SE through the availability of sustainable infrastructure by the combination of 5G networks (later 6G would be a requirement) and mixed Reality (VR+AR) to facilitate remote access to physical laboratories, the communication network must include the capability to transmit sensory data. This would allow students to experience tactile sensations such as texture, force, or weight about tangible objects within the laboratory setting to facilitate direct and immediate contact between students and physical objects, such as robots, inside laboratory settings, it is imperative for the communication network to possess the capability to transmit control data with exceptional communication performance in terms of latency, reliability, and data rates. The advent of the fifth generation of cellular communications (5G) is expected to facilitate novel prospects for future education, commonly referred to as Education 4.0. One potential game-changing technology is Ultra-Reliable Low-Latency Communications (URLLC), which facilitates the transmission of physical talents over mobile communications. The integration of Enhanced Mobile Broadband (eMBB) technology with Virtual Reality (VR) and 360◦ video streaming has the potential to offer students in virtual classrooms a very immersive educational experience. A smart campus enabled by Massive Machine Type Communications (mMTC) would enable students to access information regarding the availability of various resources, including classrooms, laboratories, and sports equipment. Additionally, it would facilitate remote booking and scheduling services for these facilities (Kizilkaya, B., et. al., 2021).

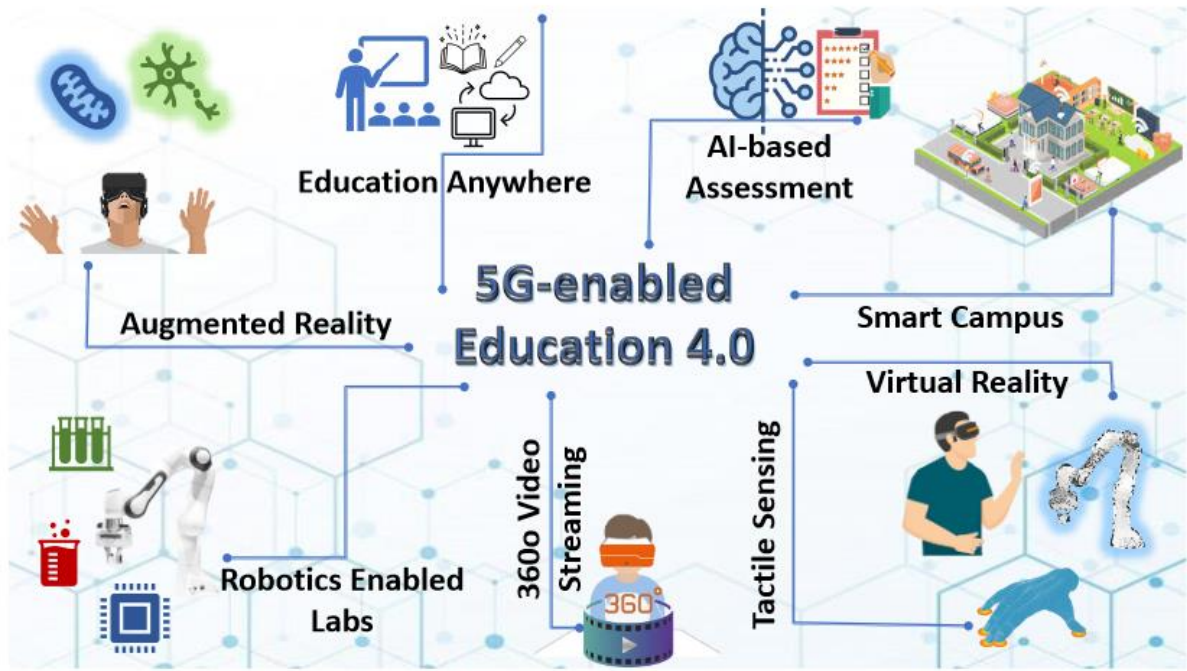

Figure 6: 5G in Education 4.0

The Business Model Canvas (BMC) is a strategic tool utilized to organize and structure the company model in a streamlined and systematic manner. The BMC is a comprehensive framework that outlines a company's value proposition, target client segments, cost structure, and revenue generation strategies. This study examines the step-by-step business model and its associated value proposition. The Value Proposition Canvas (VPC) is an effective instrument for assuring the alignment of a service or product with the needs and values of service receivers (Muhibullah, M., et al., 2021). The primary target of the author is BDA in Higher Education (HE), which BMC and VPC can evaluate. In our case, the students and staff are the service receivers whose enhanced quality and national GDP would be the reflection of the proposed BDA framework in HE.

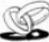

Figure 7: BMC and VPC in Education 4.0 of Bangladesh (Muhibullah, M., et al., 2021)

The efficacy of the Smart Education Framework (SEF) lies on its comprehensive description of all identified Smart Education Systems (SESs). These socioeconomic systems (SESs) range from the initial documented instance in 2010 to the most recent one in 2020. These systems predominantly exist as architectural plans, with only a limited number being partially implemented or developed as prototypes. These systems are based on various teaching and learning methodologies, including customized, individualized, adaptive, interactive, ubiquitous, collaborative, flipped, blended, case-based, and challenge-based learning. Most socio-economic statuses (SESs) utilize software that provides a wide range of functions within a learning management system. These systems integrate information technologies from multiple levels. In SESs, various information technologies are employed, including ambient intelligent classrooms, smart classrooms, virtual classrooms, interactive books, e-books, learning analytics, academic tubes, virtual reality, augmented reality, gesture-based computing, cloud computing, mobile devices, web 2.0, and social networks. Frequently utilized technology in contemporary suggestions or implementations within the socio-economic status (SES) encompasses learning analytics, electronic books (e-books), mobile devices, and cloud computing. Nevertheless, the discovered solutions did not incorporate the utilization of educational robots, serious games, or educational data mining (Demir, K.A., 2021).

There is a lack of consensus among researchers regarding a singular or universally accepted architecture for Big Data. According to several scholars, this architectural framework can be categorized into three distinct layers: the storage layer, the processing layer, and the access layer. Considering the constraints inherent in this architectural framework and with the aim of fulfilling the IoT demands of a smart campus, a proposed enhanced iteration has been put forth. This revised version is structured around six distinct levels, namely the Perception Layer, Network Layer, Storage Layer, Processing Layer, and Application Layer (Ahaidous, K., et., al., 2023).

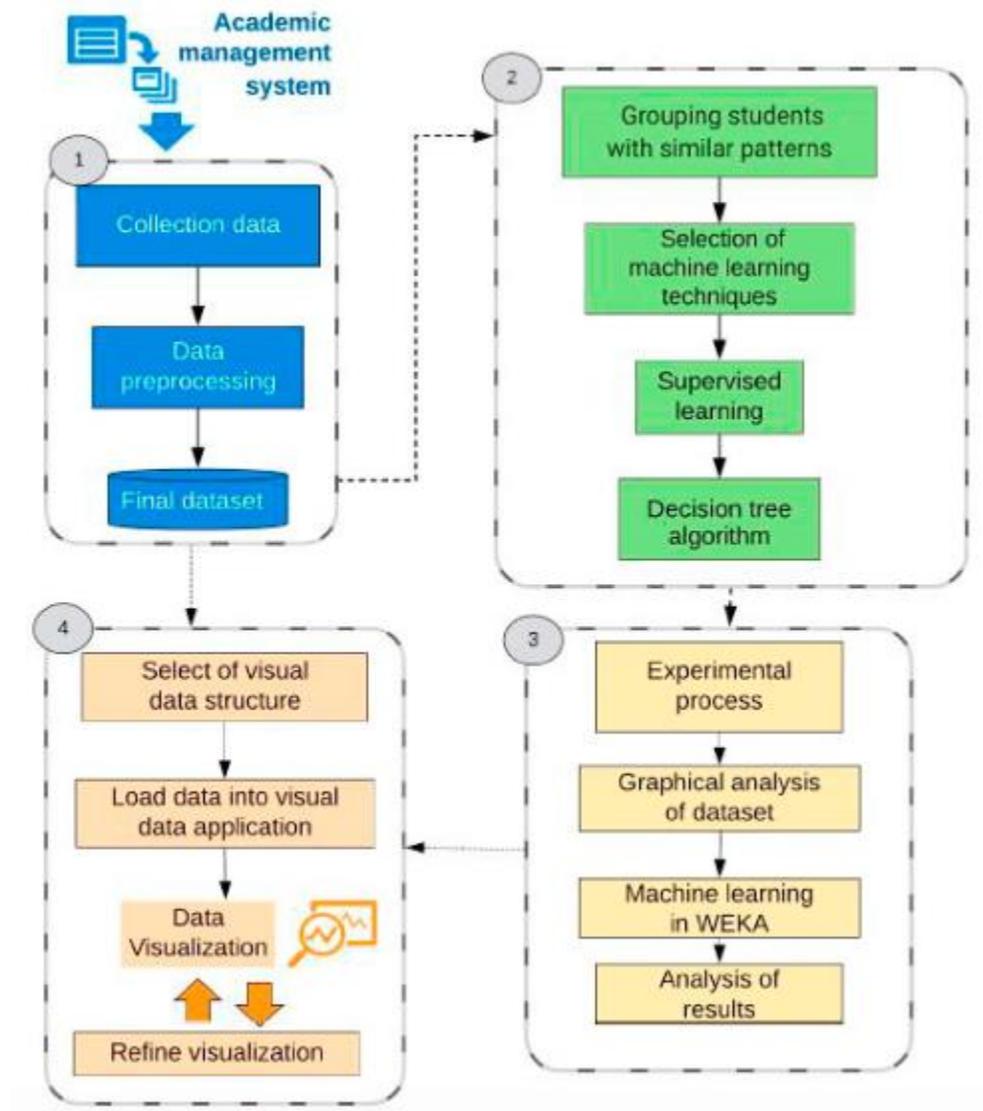

Figure 8: Big Data Architecture in Education

A latent Dirichlet allocation (LDA) topic modeling analysis was conducted, accompanied by the utilization of a visualization tool known as LDAvis. The primary objective is to conduct an analysis of the significant big data obtained from student survey replies, specifically about their perceptions of teaching behavior. Machine learning, specifically LDA topic modeling, is utilized to reveal latent regions of teaching behavior within these replies. The research demonstrates that the subjects collected possess significance and can be interpreted since they constitute coherent and semantically significant clusters of words. The analysis conducted using machine-based methods reveals the presence of eight subjects that are derived from the data. On the other hand, a manual coding analysis that follows a theory-driven approach identifies nine domains that correspond with the practical teaching behavior model, ICALT (Gencoglu, B. *et al.* 2023 ). A comparative analysis of these two methodologies reveals a convergence in frequencies, bolstering the credibility of the data-driven subjects. Novel subjects and fields of study arise, providing insights into student perspectives that extend beyond the existing framework.

Visual analytics encompasses a range of methodologies aimed at effectively managing and analyzing data, hence providing users with a multitude of viewpoints and insights. This process facilitates the identification of deviations, flaws, patterns, and noteworthy data pieces that merit additional research. The framework(Cui, Y., et al., 2021), which has been developed to offer individualized coaching to students, consists of three fundamental elements: categorization of student participants according to their collaboration abilities, analysis of the results obtained from performing an auction, and consideration of the student's chosen strategy to foster student motivation, the approach proposes the utilization of incentives that align with the unique characteristics of each learner, hence stimulating their active participation in activities that are likely to result in the attainment of these rewards. For example, students who prefer independent learning and consistently fulfill their homework obligations may accrue honor points. The BDVMFF model, if implemented, would bring about notable improvements in various aspects such as performance, efficiency, response time, computational capabilities, and grade analysis when compared to currently available Learning Management Systems (LMS), Evidence Collection (EC), Sentiment Mining (SM), Teaching Analytics (TA), and Novel Machine-Learning Technique (NMLT) approaches.

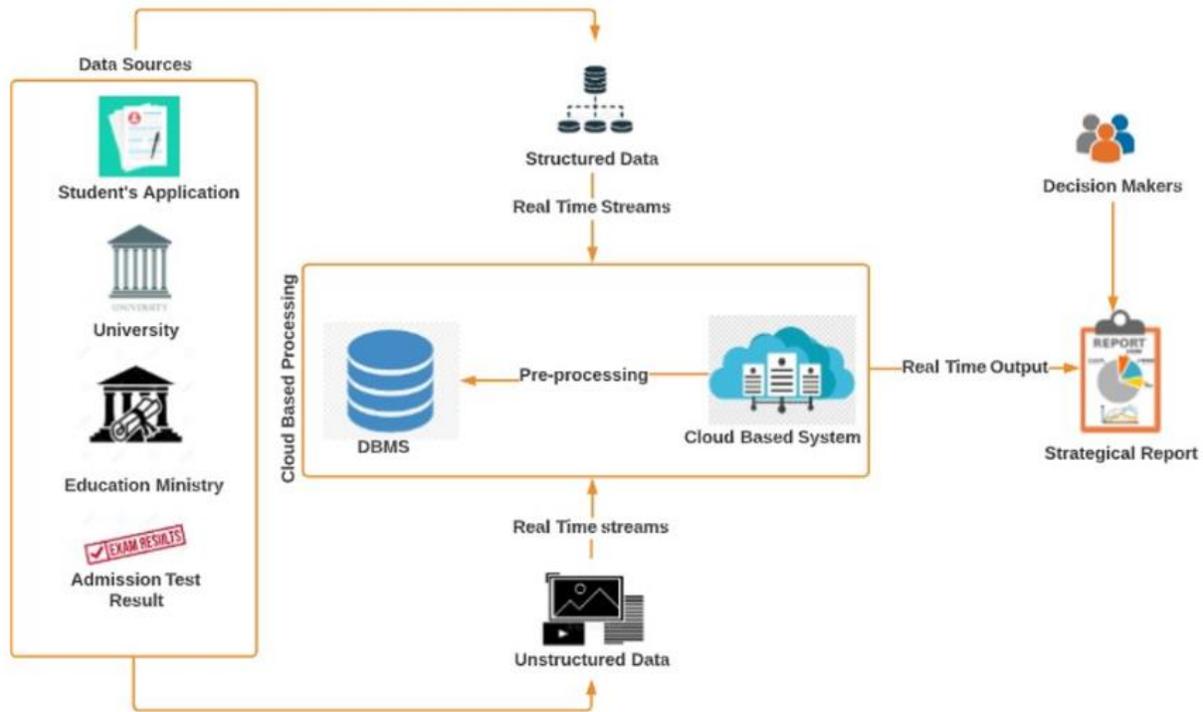

Figure 9: Big Data and Central University Admission System

A Data-Driven Decision-Making System (Johora, F.T.,et. al., 2022) is proposed to enhance Bangladesh's higher education admission process. This centralized admission system will allow students to apply for admission tests across all public and private universities. It will collect and organize admission-related data from various sources, including applications, admission results, universities, and government bodies. The system will ensure data transparency through validation and processing and will conduct data analysis to generate diverse strategic reports. Within the domain of education and big data, data mining plays a pivotal role in identifying teaching-related obstacles and the augmentation of instructors' decision-making processes, thereby elevating the overall quality of education. The region in question has implemented a specialized center for processing large volumes of data, specifically focusing on academic data, teaching evaluations, classroom observations, and employing deep learning techniques. The primary objective of this center is to develop an educational assessment and teaching platform that aims to provide comprehensive support to instructors. This platform serves as a tool for facilitating the analysis and interpretation of data, fostering a more thorough evaluation of education and teaching practices through data utilization. According to existing research, using big data technology is a valuable strategy for improving the quality of teaching and tackling various educational obstacles ( Cui, Y., et. al., 2023).

**Traditional Method + Online Tools + Big Data Analytics = Smart Education**

Risien, J., et. al. (2023) present a conceptual framework derived from collaborative, participatory methods and rigorous interpretive analysis. The text delineates how nine BID partnership teams experimented with joint structures and devised instruments to achieve their objectives internally inside their respective companies and externally in their interactions with other entities. The framework under consideration encompasses three dimensions: A, B, and C. Each of these dimensions entails the participation of many organizations, with three distinct types of players present inside each organization. The players involved in this context encompass administrators, who possess decision-making authority and are responsible for resource allocation; brokers, who operate as middlemen in enabling interactions; and delivery actors, who actively engage with the public through various activities. The paradigm largely centers on the actions and perspectives of brokers, who played a pivotal role as the most engaged participants in these collaborative ventures. It is crucial to acknowledge that the framework delineates tangible practices and should not be regarded as a prescriptive blueprint for optimal collaborations between Higher Education Institutions (HEIs) and Industry, Science, and Entrepreneurship (ISE) entities.

Extensive data integration aims to enhance results by monitoring students' actions, specifically their question-answering patterns. Customized blended learning solutions have been found to enhance classroom engagement and mitigate student attrition. The utilization of big data facilitates the examination of graduates' performance within the labor market, enabling the anticipation of potential candidates and the subsequent adaptation of recruitment efforts (Muheidat, F., et. al., 2022). The rapid increase in data availability expedites the application procedure for international students. The utilization of big data and Blockchain technology promises to transform the education industry and foster the development of more intelligent and socially beneficial individuals within society.

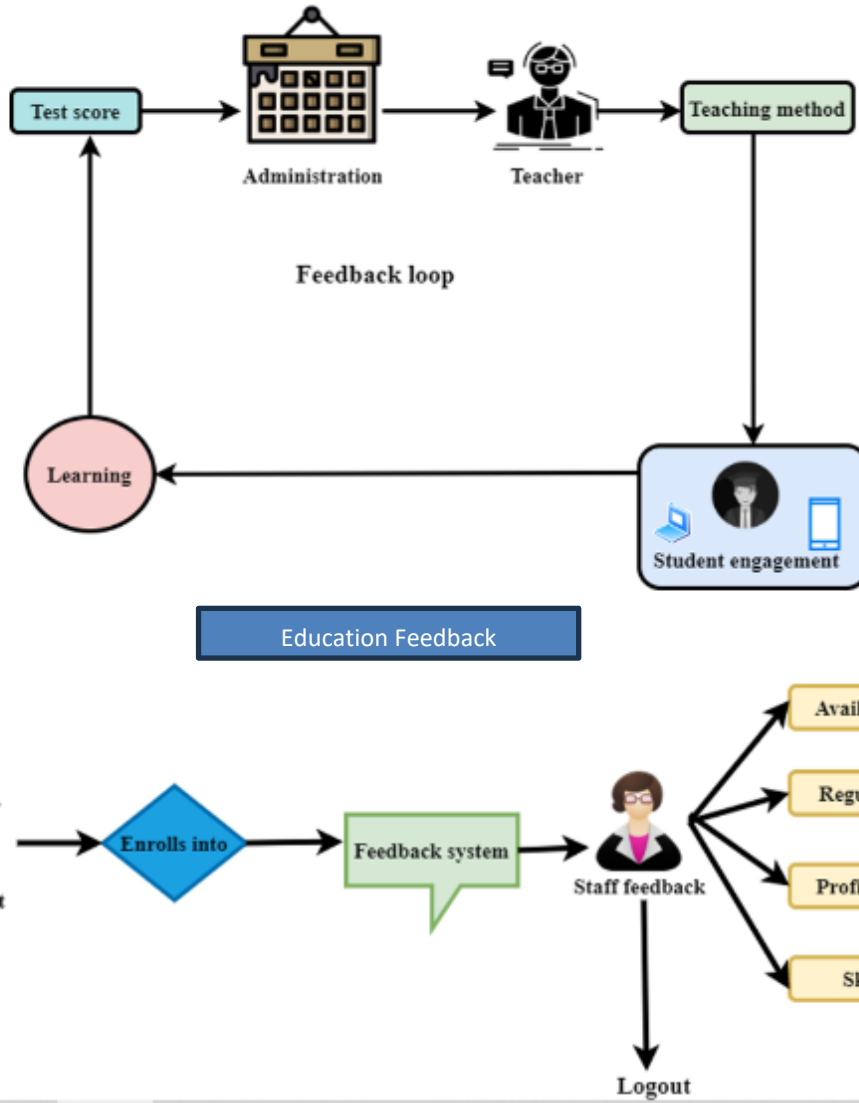

Figure 10: Data-driven Student Feedback System

Feedback on the performance of the students plays a vital role in improving their level and assists the instructor in taking appropriate measures. The Joint Information Systems Committee (JISC) in the United Kingdom has identified several instances in which universities have employed Big Data analytics to enhance educational quality and improve the experiences of both students and staff through the utilization of external data sources (A. A. Hadwer, et. al., 2019).

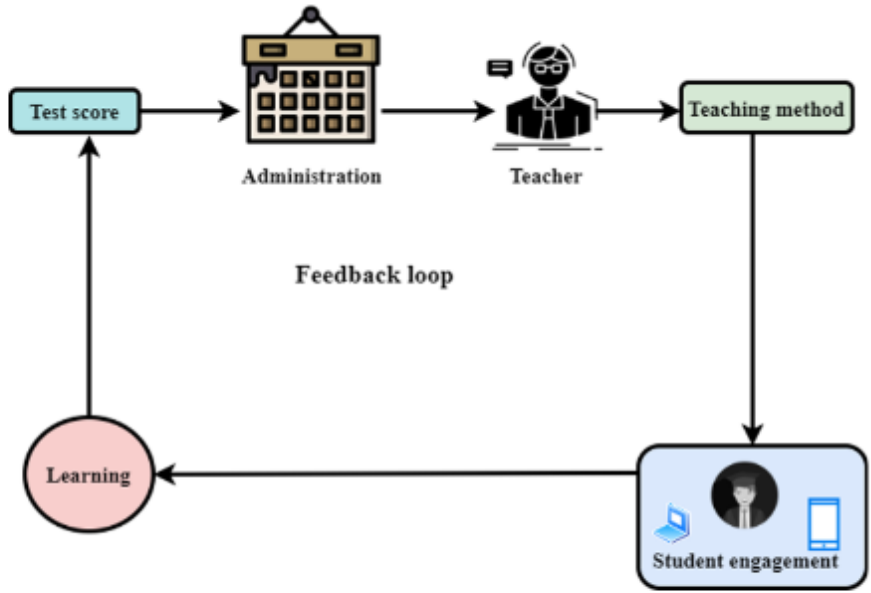

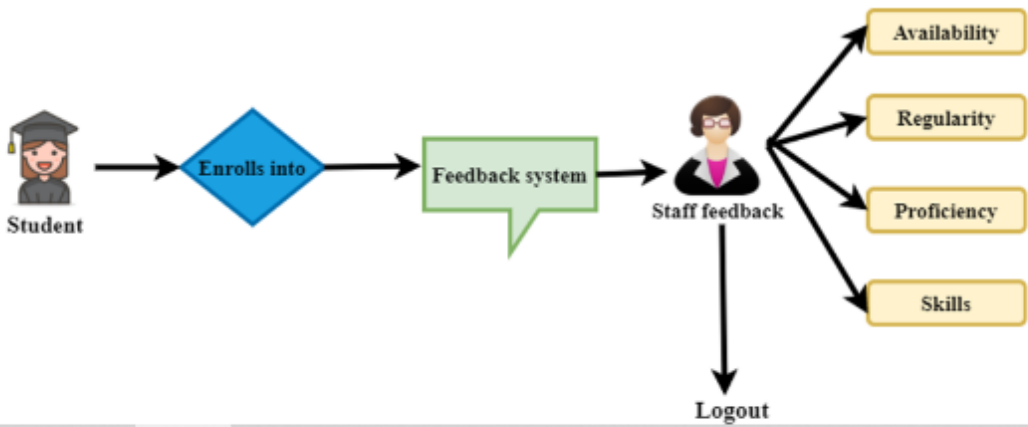

Figure 11: BDA-based feedback system for teacher and student

When the feedback is based on data, those are more reliable and authentic. A sample procedure of feedback system has been demonstrated in fig 11.

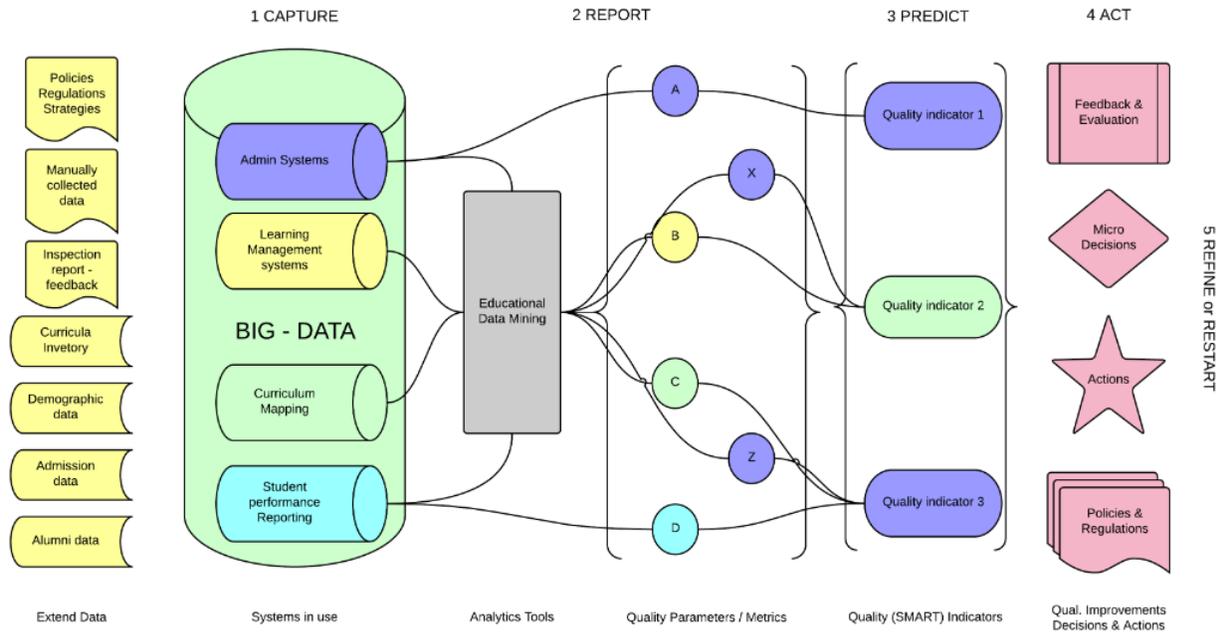

Figure 12: Framework for Quality Management in Education

Assessments within educational courses serve several purposes, encompassing the monitoring of student knowledge, facilitation of new learning, grading, and provision of summative evaluations. The objective of the study (Hervatis, V., et. al., 2015) varies among various research communities. Psychometricians place a high emphasis on the concepts of reliability and comparability, which frequently results in the implementation of standardization procedures. On the other hand, physics education research emphasizes many ideas, such as intentional practice, prompt feedback, active learning, and constructive learning. The creation of educational evaluations (Dede, C.J., et al., 2016) might vary depending on the individual educational setting and aims due to the varied objectives they aim to achieve.

## 7 Conclusion:

The utilization of Big Data has the potential to offer valuable insights in addressing the educational requirements of students. Learning analytics, as a fundamental component of Big Data in the context of higher education, gives researchers the potential to conduct real-time analysis of learning activities. Through the utilization of a retrospective study of student data, it becomes possible to develop predictive models that can be employed to investigate students who are at risk and implement suitable interventions. This, in turn, allows instructors to modify their teaching methods or provide tutoring sessions, customized assignments, and ongoing evaluation. The presence of substantial uncertainties and the continued expansion of learning analytics necessitate the examination of both the extensive possibilities for improved decision-making in higher education (Oblinger,2012) and the ethical dilemmas associated with the institutionalization of learning analytics to guide and influence student support (Slade & Prinsloo, 2013). To the best of our knowledge, the utilization of BDA in the higher education domain in Bangladesh is not noteworthy. The objective of this research was to perform an analysis aimed at identifying and developing a hierarchical structure for BDA in Bangladesh's education sector. The aim is to furnish decision-makers with a structured framework that enables them to assign priority to specific talents instead of assessing them. However, the effective utilization of BDA has the potential to facilitate significant advancements in the education sector. Instead of encountering many challenges, personalized learning environments can allow learners to tailor their educational experiences, potentially reducing the likelihood of dropouts and facilitating the development of long-term learning strategies. The utilization of Big Data Analytics can facilitate the realization of many opportunities. The opportunities can be attained through the efficient implementation and use of BDA within the higher education sector in Bangladesh. This study aims to provide a comprehensive framework for individuals from different nations to adopt comparable measures to enhance their education systems and optimize the quality of learning experiences.